# Convergent beam electron diffraction of multilayer van der Waals structures

*Tatiana Latychevskaia[1], Colin Robert Woods[2,3], Yi Bo Wang[2,3], Matthew Holwill[2,3], Eric Prestat[4], Sarah J. Haigh[2,4], Kostya S. Novoselov[2,3,5,6,7]*

[1]*Institute of Physics, Laboratory for ultrafast microscopy and electron scattering (LUMES), École Polytechnique Fédérale de Lausanne (EPFL), CH-1015 Lausanne, Switzerland*

[2]*National Graphene Institute, University of Manchester, Oxford Road, Manchester, M13 9PL, UK*

[3]*School of Physics and Astronomy, University of Manchester, Oxford Road, Manchester, M13 9PL, UK*

[4]*School of Materials, University of Manchester, Oxford Road, Manchester, M13 9PL, UK*

[5]*Department of Material Science & Engineering, National University of Singapore, 117575 Singapore*

[6]*Centre for Advanced 2D Materials and Graphene Research Centre, National University of Singapore, 117546 Singapore*

[7]*Chongqing 2D Materials Institute, Liangjiang New Area, Chongqing, 400714, China*

**ABSTRACT**

Convergent beam electron diffraction is routinely applied for studying deformation and local strain in thick crystals by matching the crystal structure to the observed intensity distributions. Recently, it has been demonstrated that CBED can be applied for imaging two-dimensional (2D) crystals where a direct reconstruction is possible and three-dimensional crystal deformations at a nanometre resolution can be retrieved. Here, we demonstrate that second-order effects allow for further information to be obtained regarding stacking arrangements between the crystals. Such effects are especially pronounced in samples consisting of multiple layers of 2D crystals. We show, using simulations and experiments, that twisted multilayer samples exhibit extra modulations of interference fringes in CBED patterns, i. e., a CBED moiré. A simple and robust method for the evaluation of the composition and the number of layers from a single-shot CBED pattern is demonstrated.

# Contents

# 1. Introduction

Convergent beam electron diffraction (CBED) has been known almost since the beginning of electron microscopy [1] and has been utilized for the study of crystallographic deformation in thick samples [2-8]. CBED offers plenty of information in a single-shot pattern. Unlike in a conventional seöected area electron diffraction pattern, in a CBED pattern each diffraction peak is turned into a finite-size CBED spot with an interference pattern-like intensity distribution that can be directly related to the three-dimensional (3D) deformations, local atomic mis-positions and strain in the crystal, as well as the sample thickness.

Two-dimensional (2D) materials have been investigated intensively by various transmission electron microscopy (TEM) techniques and have even started to be used in electron microscopy for encapsulation due to their high resilience against radiation damage [9, 10]. Recently, CBED has been demonstrated for 2D crystals and van der Waals structures where the diffraction pattern analysis is different from that for thick samples, since it is more straightforward and allows for direct structure reconstructions, such as the distance between the layers and 3D displacement of atoms [11-14]. However, it still remains a challenge to interpret the 3D structure at atomic resolution for samples consisting of more than two layers. In particular, the atomic arrangement and displacement along the $z$-direction are not trivial for reconstruction from diffraction patterns despite recent advances in imaging techniques, such as "Big Bang" tomography [15] and electron ptychography [16, 17]. Here, we present simulated CBED patterns of multilayer twisted samples and compare them with the experimentally acquired CBED patterns of multilayer van der Waals structures.

# 2. Theory and simulations

## 2.1 Transmission function

### 2.1.1 Monolayer samples

Electrons passing through a monolayer (ML) sample interact with the potential of the sample, which can be described by the following transmission function:

$$t(x, y) = \exp\left[i\sigma V_z(x, y)\right], \qquad (1)$$

where $V_z(x, y) = v_z(x, y) \otimes l(x, y)$ is the projected potential of the ML, $v_z(x, y)$ is the projected potential of a single atom, $l(x, y)$ is the function providing the positions of the atoms in the layer, $\otimes$ denotes convolution, $\sigma = \dfrac{2\pi m e \lambda}{h^2}$ is the interaction parameter, $m$ is the relativistic mass of the electron, $e$ is the elementary charge, $\lambda$ is the wavelength of the electrons, $h$ is the Planck constant, and $(x, y)$ is the coordinate in the sample plane. Phase distributions of the transmission functions for graphene and hexagonal boron nitride (hBN) MLs are shown in Fig. 1 (details regarding the simulation are provided in Appendix A). In graphene, a single carbon atom causes a phase shift up to 0.217 radian while the graphene ML causes a phase shift up to 0.221 radian. Single B and N atoms cause phase shifts up to 0.190 and 0.238 radian, respectively, and the hBN ML causes a phase shift up to 0.245 radian. Both graphene and hBN MLs can be considered as weak phase objects for typical accelerating voltages used in a TEM.

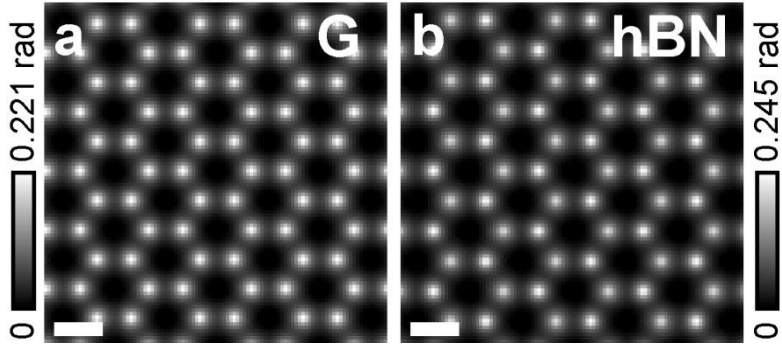

Fig. 1. Phase distributions of transmission functions of graphene (a) and hBN (b) monolayers. The scalebars are 2 Å.

### 2.1.2 Bilayer samples

For twisted bilayer graphene (TBG), the approximation of a weak phase object also holds, and when neglecting the diffraction effects due to propagation between the two layers, the TBG sample can be assigned the following transmission function:

$$
\begin{aligned}
t_{\text{TBG}}(x,y) &= \exp\left[i\sigma V_z^{(1)}(x,y)\right]\exp\left[i\sigma V_z^{(2)}(x,y)\right] \approx \\
&\approx \left[1 - i\sigma V_z^{(1)}(x,y)\right]\left[1 - i\sigma V_z^{(2)}(x,y)\right] = \\
&= 1 - i\sigma V_z^{(1)}(x,y) - i\sigma V_z^{(2)}(x,y) - \sigma^2 V_z^{(1)}(x,y)V_z^{(2)}(x,y),
\end{aligned}
\tag{2}
$$

where $V_z^{(1)}(x,y)$ and $V_z^{(2)}(x,y)$ are the projected potentials of layers 1 and 2, respectively. The last term in Eq. 2 describes the overlap between the two lattice's potentials, which leads to the formation of the moiré structure. TBG imaged using TEM in diffraction mode exhibits a set of diffraction peaks, corresponding to each individual layer. The diffraction peaks corresponding to the moiré structure are not observed at typical TEM electron energies (30 – 300 keV) due to a low value of the interaction parameter $\sigma$ [18, 19], $\sigma \approx 0.002 - 0.001$ 1/VÅ for 30 – 300 keV, respectively. Recently, diffraction peaks due to the moiré structure were observed in diffraction patterns acquired with low-energy electrons of 236 eV [20], where the interaction parameter $\sigma$ is relatively large, $\sigma \approx 0.02$ 1/VÅ.

### 2.1.3 Multilayer samples

The transmission function of a multilayer sample, where all MLs are in the exact same stacking with the same relative rotation, neglecting the propagation between the layers, can be written as:

$$t(x,y) = \exp[iN_i \sigma V_z(x,y)], \quad (3)$$

where $N_i$ is the number of layers. The phase shift introduced by a set of layers is given by the phase shift of a single layer multiplied with the number of layers. When the number of layers is five or more, the total phase shift exceeds 1 rad and such samples cannot be considered as weak phase objects.

We consider a situation where the multiple layers are arranged into two sets of layers, each set is made up of several layers with the exact same stacking and the same relative rotation, as illustrated in Fig. 2. A lattice mismatch or a twist rotation between sets 1 and 2 can give rise to a moiré structure. In general, for such a multilayer sample the approximation of a weak object is not fulfilled.

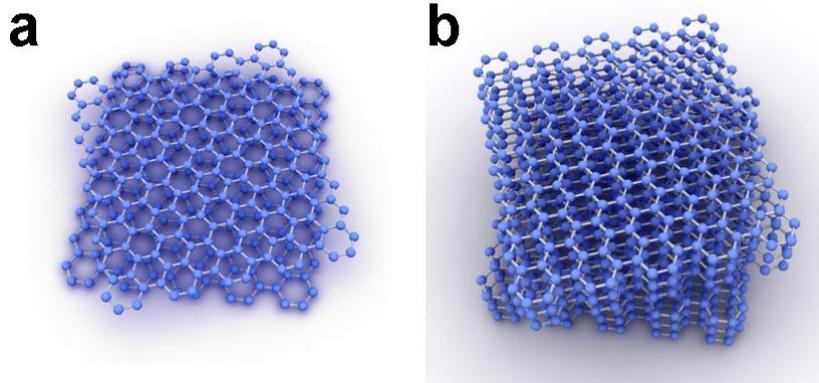

Fig. 2. Illustration of multilayer graphene: Bilayer ($N_1 = N_2 = 1$) and a few layer graphene ($N_1 = N_2 = 3$).

### 2.2 Diffraction patterns

The diffraction patterns of a twisted graphene multilayer sample simulated at 80 keV are shown in Fig. 3. Neglecting the propagation between the layers, the transmission function of the entire sample was assumed as a product of the two transmission functions corresponding to each set:

$$t(x,y) = t_1(x,y) t_2(x,y) = \exp[i\sigma N_1 V_z^{(1)}(x,y)] \exp[i\sigma N_2 V_z^{(2)}(x,y)], \quad (3)$$

where $N_1$ and $N_2$ are the number of layers in each set, and $V_z^{(1)}(x,y)$ and $V_z^{(2)}(x,y)$ are the projected potentials of each set. The diffraction patterns were calculated as the square of the amplitude of the Fourier transform of $t(x, y)$, where the Fourier transform was calculated by FFT. No weak phase object approximation was applied in the simulations.

Figure 3(a) shows the simulated diffraction pattern of a multilayer graphene sample consisting of two sets of graphene layers, with each set consisting of ten layers, and the relative twist between the sets is 10°. The peaks due the moiré structure are apparent in the diffraction pattern. Figure 3(b) shows the plots of the intensity and the ratio of the first-order peaks and the peaks resulting from the moiré structure as a function of the number of layers in each set. The number of layers is the same in each set and it ranges from 1 to 20 layers. When the number of layers exceeds around ten layers, the ratio between the peaks due to the moiré structure the first-order peaks and becomes 0.001 and the moiré peaks can be detected in the diffraction pattern.

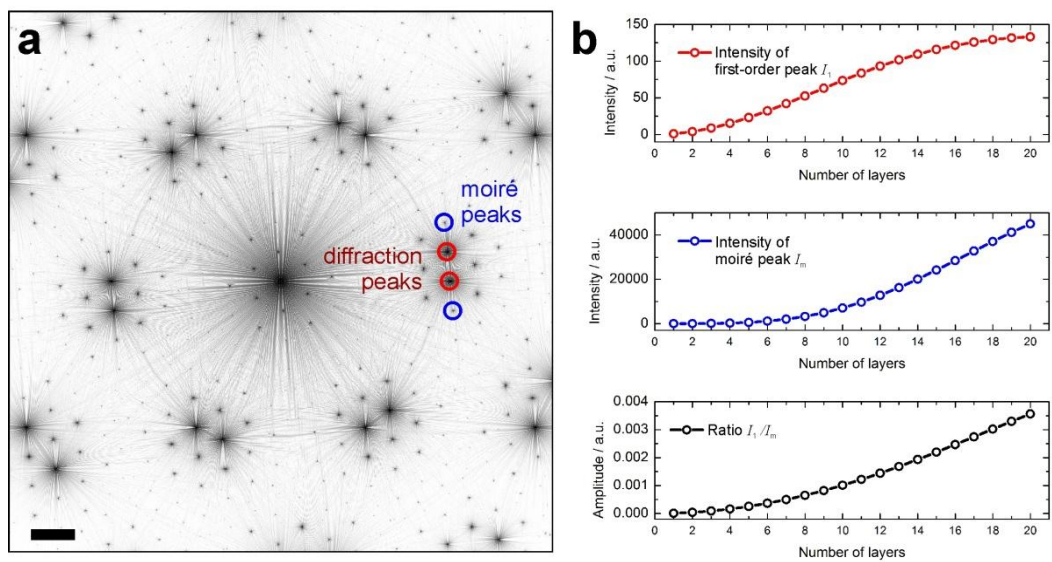

Fig. 3. Simulated electron diffraction patterns of a multilayer graphene sample consisting of two sets of graphene layers. The relative rotation between the sets is 10°. (a) Diffraction pattern of two set of graphene layers where each set consists of ten layers. The first order peaks are indicated with the red circles. The peaks due to diffraction on the moiré structure are indicated with the blue circles. The scalebar is 2 nm$^{-1}$. (b) Intensity of the first-order and moiré peaks and the ratio of the intensity of the moiré peaks to the first-order peaks as a function of the number of layers in each set. The number of layers is the same in each set.

The simulations show that for a small number of layers, up to six layers, the intensity of the first-order diffraction peaks exhibits approximately quadratic dependency on the number of layers, while the intensity of the moiré peaks exhibits approximately 4$^{th}$ degree polynomial dependency on the number of layers. According to these simulations, the moiré peaks have sufficient intensity to be detected when the number of layers exceeds approximately ten layers.

## 2.3 CBED interference and moiré

The CBED experimental arrangement is sketched in Fig. 4. The CBED patterns of multilayer samples simulated at 80 keV are shown in Fig. 5 and the simulation procedure is explained in Appendix B. Here, the samples consist of sets of layers, with each set consisting of the same type (graphene or hBN) of layers that are rotated by the same twist angle. The simulated CBED patterns exhibit characteristic six-fold-symmetry arrangement of CBED spots. Each CBED spot exhibits sharp edges, because the intensity distribution is given by the image of the limiting aperture.

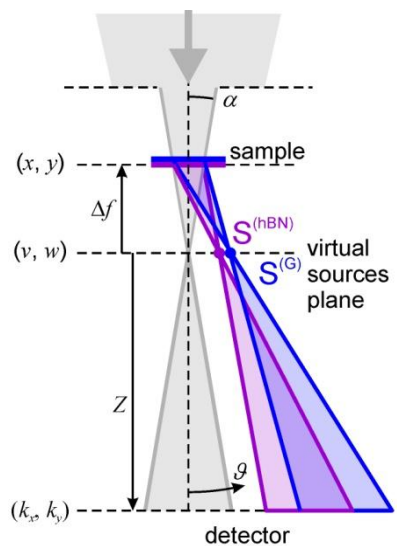

Fig. 4. Sketch of CBED experimental arrangement.

A simulated CBED pattern of a bilayer (BL) graphene-hBN sample is shown in Fig. 5(a). Here, the intensity distribution in a selected CBED spot resembles an interference pattern created by two point sources. The positions of the virtual sources in the virtual source plane are the same as the positions of the diffraction peaks (similar to those shown in the red circles in Fig. 3(a)) in the corresponding diffraction pattern, as sketched in Fig. 4. We will refer to such an interference pattern as "CBED interference", as indicated in Fig. 5(b). When the number of layers in either set (graphene or hBN) increases, then additional modulation along the CBED interference fringes begins to emerge, as shown in Figs. 5(c) – (f). For a large number of layers in both sets, such modulations become even more pronounced, in particular in the higher-order CBED spots. This is illustrated in Figs. 5(e) and (f), where a simulated CBED pattern of a sample consisting of ten layers of graphene and ten layers of hBN is shown. One can trace the emergence of these extra modulations as coming from the interference between the moiré CBED spots, similar to the moiré spots observed in the diffraction patterns (indicated by the blue circles in Fig. 3(a)). We therefore will refer to these intensity modulations as "CBED moiré" as indicated in Fig. 5(f). The moiré CBED spots originating from the virtual sources corresponding to the moiré peaks are not directly visible in the CBED patterns due to their weak intensity and because moiré CBED spots strongly overlap with the major CBED spots. However, moiré CBED spots manifest themselves in the CBED moiré, which is created by the interference between the moiré CBED spots and the major CBED spots.

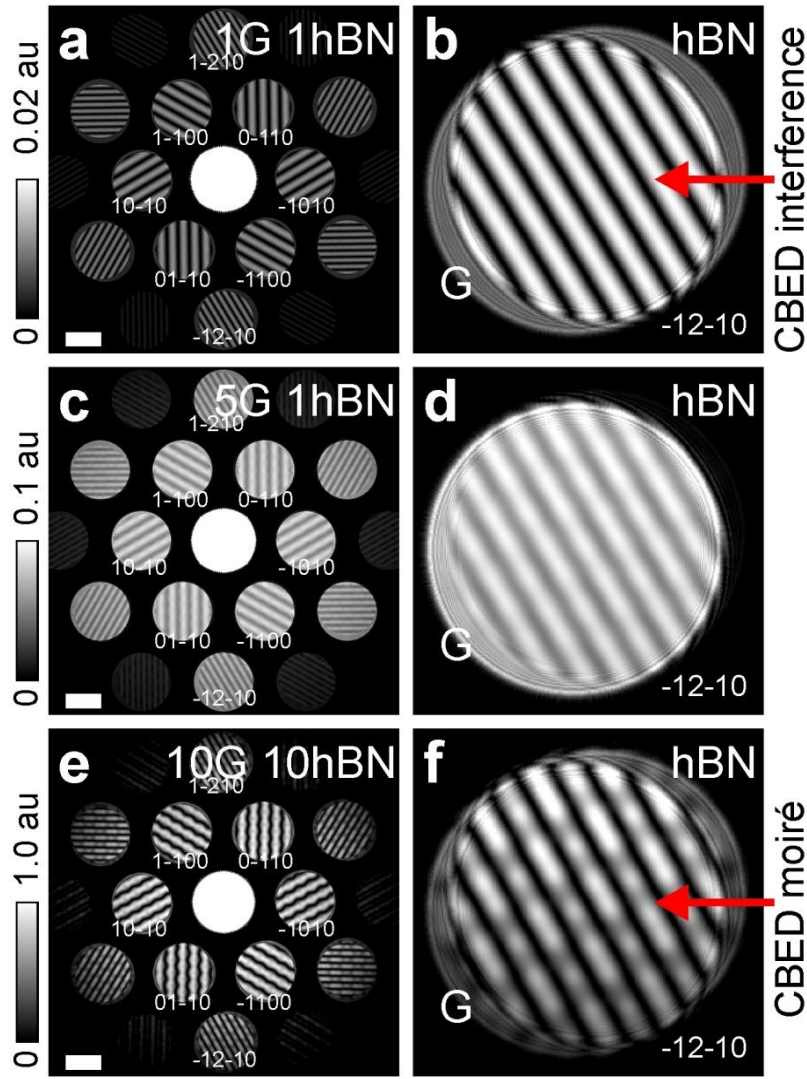

Fig. 5. Simulated convergent beam electron diffraction (CBED) of multilayer van der Waals structures at a defocus, $\Delta f$ = 2.0 µm, the twist angle between graphene and hBN layers is 2°. (a) Bilayer system of graphene and hBN, and (b) magnified CBED spot (-12-10) where CBED interference fringes are observed. (c) System of five layers of graphene and one layer of hBN and (d) magnified CBED spot (-12-10). (e) Ten layers of graphene and ten layers of hBN, and (f) magnified CBED spot (-12-10) where a CBED moiré is observed in CBED interference fringes. The scalebars in (a), (c) and (e) are 2 nm$^{-1}$.

## 2.4 Relation between diffraction and CBED patterns

The CBED moiré can be explained by comparing the diffraction and CBED patterns of the twisted multilayer sample, as illustrated in Fig. 6. Figures 6(a) and (b) show the simulated diffraction pattern of sample consisting of one layer of graphene and one layer of hBN with a twist angle of 2°. Two sets of intense six-fold arranged diffraction peaks are formed by the diffraction on the individual lattices. The satellite peaks are the moiré peaks, which are formed by the diffraction on the moiré structure. These moiré peaks are less intense and become noticeable only when the number of layers is ten or more, as shown in Figs. 6(e), (f), (i) and (j).

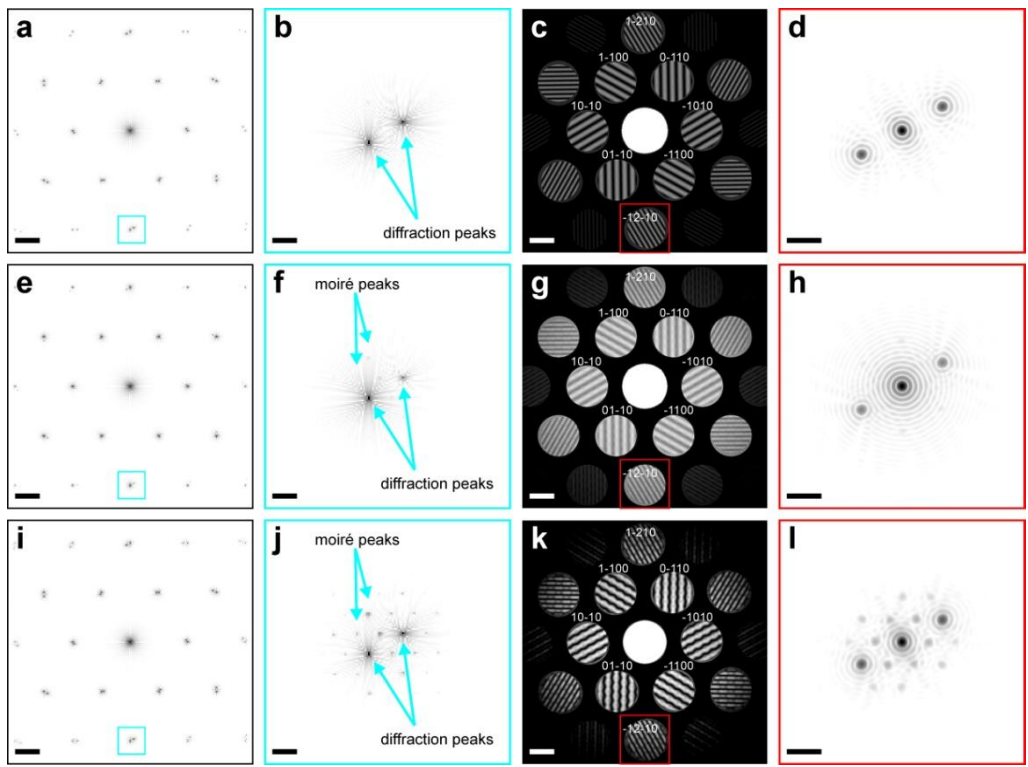

Fig. 6. Simulated diffraction and CBED patterns of sample consisting of graphene and hBN layers with twist angle of 2°. (a) – (d) One graphene and one hBN layers, (e) – (h) five graphene and one hBN layers and (i) – (l) ten graphene and ten hBN layers. (a), (e) and (f) Diffraction patters. (b), (f) and (j) Magnified regions in the cyan square in (a), (e) and (f), respectively. (c), (g) and (k) CBED patterns at $\Delta f$ = 2.0 μm. (d), (h) and (l) Amplitude of the Fourier transform of spots (-12-10) in the red squares in (c), (g) and (k), respectively. The scalebars are: (a), (c), (e), (g), (i) and (k) 2 nm$^{-1}$; (b), (f) and (j) 0.2 nm$^{-1}$; (d), (h) and (l) 2nm. (a), (b), (d), (e), (f), (h), (i), (j) and (l) are shown in the inverted intensity scale.

Figures 6(c) and (d) show the simulated CBED pattern of the same sample as in Figs. 6(a) and (b). The fringed interference pattern in a CBED spot can be represented as a far-field interference pattern created by waves originating from virtual sources. The intensity of each virtual source is given by the intensity of the corresponding diffraction spot, which in turn is given by the number of layers.

In the case of the TBL sample, there are only two sets of intense diffraction peaks, Figure 6(a) and (b). For two virtual sources, the intensity distribution in a CBED spot can be described as:

$$I(k_x, k_y) = |a_1|^2 + |a_2|^2 + 2a_1 a_2 \cos(k_x \Delta v + k_y \Delta w), \qquad (1)$$

where $(\Delta v, \Delta w)$ is the distance between the two virtual sources, $a_1$ and $a_2$ are the amplitudes of the emitted waves, and $(k_x, k_y)$ is the coordinate in the detector plane. A Fourier spectrum of a selected CBED spot, obtained by calculating the Fourier transform of the intensity distribution in the selected spot, provides indirect information regarding the distribution of the virtual sources, Figs. 6(c) and (d). The positions of the virtual sources can be estimated from the positions of the peaks in the Fourier spectra. The amplitudes $a_1$ and $a_2$ can be evaluated from the amplitudes of the peaks in the Fourier spectra. These amplitudes can then be related to the number of layers.

Figures 6(e) and (f) show the simulated diffraction pattern of a sample consisting of five graphene layers and one hBN layers with a relative twist 2°. Figure 6(g) shows the corresponding CBED pattern. Here, two sets of intense six-fold arranged diffraction peaks are observed in the diffraction pattern (Figs. 6(e) and (f)), and the corresponding peaks are observed in the spectra of an individual CBED spots (Figs. 6(g) and (h)). Moiré peaks are not observed in the diffraction pattern (Fig. 6(e)), the CBED moiré is not observed in the CBED pattern (Fig. 6(g)) and the spectra of a selected CBED spot exhibit very weak indication of the peaks related to the CBED moiré (Fig. 6(h)).

In the case of the multilayer sample, there are moiré peaks in addition to the intense diffraction peaks in the diffraction patterns, as shown Figs. 6(i) and (j). In addition, the CBED moiré is clearly seen as intensity modulations in CBED interference patterns, Fig. 6(k). The spectra of a selected CBED spot exhibit peaks related to the CBED moiré (Fig. 6(l)).

## 3. Estimation of number of layers from a single CBED pattern

Many methods allow for the determination of the number of layers in a multilayer sample at a spatial resolution from tens of nanometres to several microns - by an optical contrast [21], atomic force microscopy [22], Raman spectroscopy [23, 24] electron energy loss spectroscopy [25, 26], or by tilting the sample and measuring it in electron diffraction mode [27]. Simulations have shown that dark-field scanning transmission electron microscopy can be applied to determine the number of layers on a scale of atomic distances [28]. For a bilayer samples, the averaged interlayer distance at sub-Ångstrom precision can be obtained by tilting the sample in selected area electron diffraction measurements and fitting the intensity maxima with the theoretical model [29]. Recently, it was shown that the interlayer distance in bilayer samples can be measured at nanometre spatial resolution and sub-Ångstrom precision from a single image by CBED [13]. However, simultaneous measurement of the number of layers and the separation distance between the layers in multilayer samples still remains a challenge. The most successful approach, cross-sectional TEM imaging [30], requires sophisticated sample preparation and is a destructive procedure that takes several hours. A technique that would allow us to evaluate the number of layers and the distance between the layers would be a valuable tool for the characterization of 2D materials. Here, we propose a method that allows quick estimation of the number and composition of layers from a single CBED pattern.

The composition and the relative number of layers in a multilayer sample can be evaluated from the intensity at the rims of a selected CBED spot, in the regions where the CBED spots from different types of layers do not overlap, as shown in Fig. 5(b). CBED spots (as well as diffraction peaks) originating from a lattice with a smaller period are found at higher scattering angles, allowing us to assign CBED spots to graphene and hBN layers. Simulations show that the intensity of a first-order CBED spot of graphene ML is 1.12 times higher than that of the hBN monolayer. Thus, the relative number of layers can be evaluated as follows. In a selected CBED spot, the overlapping spots are assigned to graphene or hBN based on their radial position (graphene CBED spots are positioned radially further from the centre of the CBED pattern). Two crescent-shaped areas at the CBED spot opposite sides that belong to different layers are selected. For each selected area, the averaged intensity (also averaged over the six CBED spots of the same order) is calculated and the intensity ratio graphene/hBN is obtained. For the simulated CBED pattern of the graphene-hBN bilayer sample shown in Figs. 5(a) and (b), the intensity ratio estimated from a first-order CBED spot from the edges of the spot is 1.15.

The absolute number of layers is more difficult to evaluate precisely, but an estimation can be done from the Fourier spectrum of the selected spot, where samples with five and more layers exhibit peaks due to the CBED moiré. For a multilayer sample, diffraction and moiré peaks lead to multiple virtual sources in the virtual source plane and as a result, the total interference pattern in a CBED spot is described by superposition of multiple wavefronts. It is in principle possible to retrieve the amplitudes and the positions of the individual virtual sources from a given interference pattern, but this is not a trivial task. In contrast, a simple Fourier transform of a CBED spot intensity distribution already provides plethora of information. For example, the presence of moiré peaks already indicates that the system consists of more than just two layers. A priori knowledge about the sample can help to narrow down this information to a precise number of layers, as we show below in an experimental example.

## 4. Experimental

The samples were prepared by a pick and lift method. Graphene was exfoliated onto a PMMA substrate and the BN onto a SiO$_2$ substrate. Then, the PMMA was used as a membrane to suspend the graphene over the hBN crystal as the two are brought into contact. The crystals adhere and the membrane was lifted again with the two crystals attached. The two crystals were then positioned over a hole in the TEM grid and brought into contact with it. The PMMA was then removed with acetone [31]. TEM CBED imaging was performed with a probe side aberration corrected Titan ChemiSTEM operated at 80 kV with a small convergence angle and a probe current of 110 pA. The images were recorded with a 16 bit intensity dynamic range detecting system, without using a beam spot, so that the intensity in all CBED diffraction spots is available. Each image was the average of ten identical acquisitions, with a 1 s acquisition time.

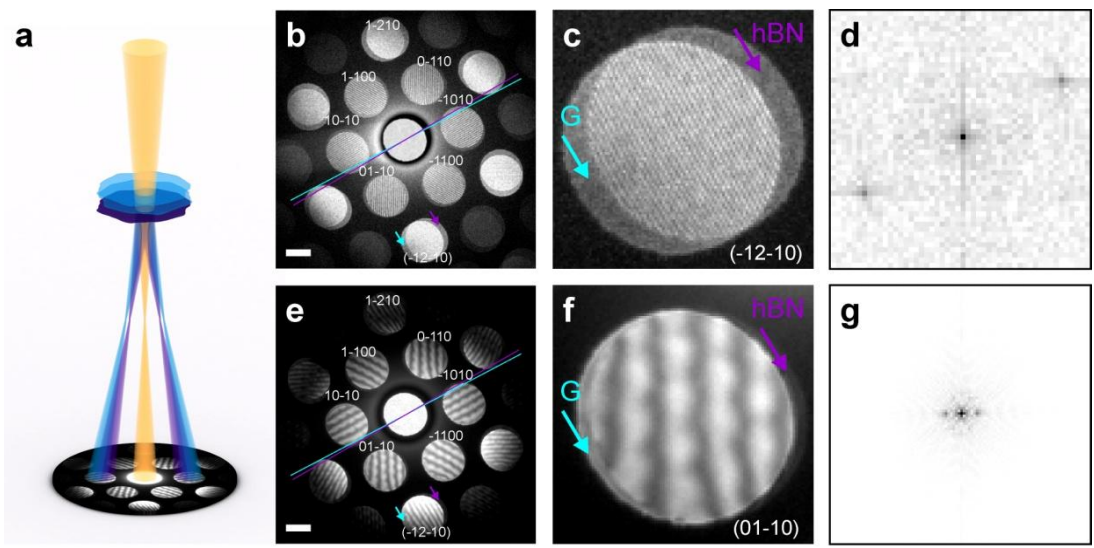

Fig. 7. Experimental CBED of multilayer van der Waals structures. (a) Schematics of the experimental arrangement. (b) CBED patterns acquired at $\Delta f$ = 6.0 µm, (c) magnified image of (-12-10) CBED spot and (d) amplitude of its Fourier transform. (e) CBED patterns acquired at $\Delta f$ = 2.0 µm, (f) magnified image of (10-10) CBED spot and (g) amplitude of its Fourier transform. The scalebars in (b) and (e) are 2 nm$^{-1}$.

The experimental scheme is depicted in Fig. 7(a) and the acquired CBED patterns are shown in Figs. 7(b) and (e). The CBED pattern acquired at a defocus, $\Delta f$ = 6.0 µm (shown in Fig. 7(b)) exhibits two sets of CBED spots of almost same intensity, as indicated by the cyan and the lilac arrows in Figs. 7(b) and (c). This implies that the number of graphene layers is the same as the number of hBN layers. The intensity ratio for non-overlapping areas was calculated to be G/hBN=1.07, and the relative number of layers of G/hBN=0.94≈1 was obtained. The Fourier spectrum of a selected spot exhibits no peaks due to the CBED moiré (Fig. 7(d)). The absence of CBED moiré implies that the sample most probably was a BL sample.

Another example CBED pattern of the graphene-hBN sample is shown in Figs. 7(e) and (f). Here, the CBED pattern acquired at $\Delta f$ = 2.0 μm exhibits two sets of CBED spots of different intensity. The set of spots corresponding to graphene has noticeably higher intensity than the set of spots corresponding to hBN, as indicated by the cyan and the lilac arrow in Figs. 7(e) and (f). This implies that there are more graphene layers than hBN layers in the sample. To evaluate the relative number of layers, the average intensity was calculated at the non-overlapping areas as described above and the intensity ratio G/hBN=2.72 was obtained. Taking into account that intensity of CBED of graphene layer is 1.15 higher than that of hBN, we obtain the relative number of layers of G/hBN=2.36. The Fourier spectrum of a selected spot exhibits weak peaks due to CBED moiré (Fig. 7(g)). The presence of the CBED moiré implies that the sample most probably was a multilayer sample with five graphene and two hBN layers.

## 5. Discussion and conclusions

We investigated the CBED imaging of samples consisting of multiple layers of 2D crystals by simulations and experiments. We showed that twisted multilayer samples, unlike BL samples, exhibit a CBED moiré, extra modulations of interference fringes in CBED spots. The composition and the relative number of layers can be evaluated from the intensity distribution in the non-overlapping regions of a CBED spot. A more precise estimation of the number of layers can be done from a Fourier spectrum of a CBED spot, where the presence of peaks due to the CBED moiré indicate that there are five or more layers in the sample. Although the precision of the sample characterisation with this technique is very modest when compared to cross-sectional TEM imaging, the presented approach has the advantage that is non destructive, it requires only a single shot CBED pattern, and CBED is relatively easy to realise in a conventional TEM.

# Appendices

## Appendix A: Transmission function of monolayer

The transmission functions of MLs were calculated as follows. The transmission function of a ML can be written as:

$$t(x,y) = \exp[i\sigma V_z(x,y)] = \exp[i\sigma v_z(x,y) \otimes l(x,y)], \quad (A1)$$

where $v_z(x,y)$ is the projected potential of an individual atom, $l(x,y)$ is the function describing positions of the atoms in the lattice, and $\otimes$ denotes convolution. The projected potential of a single carbon atom was simulated in the form [32]:

$$v_z(r) = 4\pi^2 a_0 e \sum_i a_i K_0\left(2\pi r\sqrt{b_i}\right) + 2\pi a_0 e \sum_i \frac{c_i}{d_i} \exp(-\pi^2 r^2 / d_i),$$

where $r = \sqrt{x^2 + y^2}$, $a_0$ is the Bohr' radius, $e$ is the elementary charge, $K_0(...)$ is the modified Bessel function, and $a_i, b_i, c_i, d_i$ are parameters that depend on the chemical origin of the atoms and are tabulated in Ref. [32]. The analytical expression for $v_z(r)$ has singularity at $r = 0$, but because an atom has a finite size with the radius of approximately $r$ = 0.1 Å, $v_z(r)$ at $r$ = 0 was replaced by the value at $r$ = 0.1 Å. The convolution $v_z(x,y) \otimes l(x,y)$ in Eq. A1 was calculated as $\mathrm{FT}^{-1}\{\mathrm{FT}[v_z(x,y)]\mathrm{FT}[l(x,y)]\}$, where FT denotes Fourier transform. $\mathrm{FT}[l(x,y)]$ was simulated as $\mathrm{FT}[l(x,y)] = \sum_n \exp[-i(k_x x_n + k_y y_n)]$, where $(x_n, y_n)$ are the atomic positions, without applying Fast Fourier transforms (FFT) to avoid artifacts associated with FFT. The distributions in the sample plane were sampled with 1 pixel = 0.142 × 0.142 Å$^2$, and 5634 × 5634 pixels, which gives the sample size of 80 × 80 nm$^2$. This gives the pixel size in the diffraction plane $\Delta k$ = 1.25·10$^7$ m$^{-1}$. The inverse Fourier transform was calculated by applying inverse FFT to the product of $\mathrm{FT}[v_z(x,y)]$ and $\mathrm{FT}[l(x,y)]$.

## Appendix B: Simulated CBED patterns of twisted multilayer sample

CBED patterns were simulated as follows. The incident wavefront distribution $\psi_0(\vec{r})$ was calculated by simulation diffraction of the spherical wavefront on a limiting aperture (second condenser aperture) positioned at a plane $\vec{r}_0$:

$$\psi_0(\vec{r}) \propto \iint a(\vec{r}_0) \frac{\exp(-ikr_0)}{r_0} \frac{\exp(ik|\vec{r}_0 - \vec{r}|)}{|\vec{r}_0 - \vec{r}|} d\vec{r}_0, \quad (B1)$$

where $a(\vec{r}_0)$ is the aperture function. Each ML was assigned a transmission function $t_i(x_i, y_i)$ defined by Eq. 1, where $i = 1, 2 ... M$ is the layer number. No weak phase object approximation was applied in the simulations. The exit wave after passing through the first layer was given by $u_1(x_1, y_1) = \psi_0(x_1, y_1) t_1(x_1, y_1)$. Next, this wave was propagated to the second layer. The propagation was calculated by the angular spectrum method [32-34]. The propagated wave was described by the complex-valued distribution $u_{2,0}(x_2, y_2)$. The exit wave after passing through the second ML was calculated as $u_2(x_2, y_2) = u_{2,0}(x_2, y_2) t_2(x_2, y_2)$ and so on, the electron wave propagation through all the layers is calculated. The CBED was then simulated as the square of the amplitude of the Fourier transform of $u_M(x_M, y_M)$, where the Fourier transform was calculated by FFT.

## Acknowledgements

This work was supported by the European Union Graphene Flagship Program, European Research Council Synergy Grant 319277 "Hetero2D" and European Research Council Starting Grant 715502 "EvoluTEM"; the Royal Society; Engineering and Physical Research Council (UK); and US Army Research Office (grant W911NF-16-1-0279). S.J.H and E.P. acknowledge funding from the Defence Threat Reduction Agency (grant HDTRA1-12-1-0013) and the Engineering and Physical Sciences Research Council (UK) (grants EP/K016946/1, EP/L01548X/1, EP/M010619/1 and EP/P009050/1).